\documentclass[twocolumn]{aastex631}

\usepackage{amsmath}
\usepackage{amssymb}
\usepackage{graphicx}
\usepackage{hyperref}

\usepackage{xspace}

\newcommand{\Fg}[1]{Figure~\ref{fig:#1}}%beginning of the sentence

\newcommand{\eq}[1]{Eq.~(\ref{eq:#1})\xspace}
\newcommand{\Se}[1]{Section~\ref{sec:#1}\xspace}%beginning of the sentence

\newcommand{\ie}{i.e.}

\newcommand{\eg}{e.g.}

\newcommand{\Kpc}{ \  \rm kpc}
\newcommand{\Mpc}{ \  \rm Mpc}

\newcommand{\Mstar}{ \  \rm M_{\star}}

\newcommand{\Msun}{ \   \rm M_{\odot}}

\graphicspath{{./}{figures/}}
\begin{document}

\title{A rotating satellite plane around Milky Way-like galaxy from the TNG50 simulation}

%% The \author command is the same as before except it now takes an optional
%% argument which is the 16 digit ORCID. The syntax is:
%% \author[xxxx-xxxx-xxxx-xxxx]{Author Name}
%%
%% This will hyperlink the author name to the author's ORCID page. Note that
%% during compilation, LaTeX will do some limited checking of the format of
%% the ID to make sure it is valid. If the "orcid-ID.png" image file is 
%% present or in the LaTeX pathway, the OrcID icon will appear next to
%% the authors name.
%%

\correspondingauthor{Yingzhong Xu, Xi Kang}
\email{xuyingzhong@zju.edu.cn, kangxi@zju.edu.cn}

\author{Yingzhong Xu}
\affiliation{Institute for Astronomy\\
the School of Physics, Zhejiang University\\
38 Zheda Road, Hangzhou 310027, China}

\author{Xi Kang}
\affiliation{Institute for Astronomy\\
the School of Physics, Zhejiang University\\
38 Zheda Road, Hangzhou 310027, China}
\affiliation{Purple Mountain Observatory\\
10 Yuan Hua Road, Nanjing 210034, China}

\author{Noam I. Libeskind}
\affiliation{Leibniz-Institut f\"ur Astrophysik Potsdam (AIP)\\
An der Sternwarte 16, D-14482 Potsdam, Germany}

\begin{abstract}

We study the Satellite Plane Problem of the Milky Way\ (MW) by using the recently published simulation data of TNG50-1. Here, we only consider the satellite plane consisting of the brightest 14 MW satellites \ (11 classical satellites plus Canes Venatici I\ (CVn
I), Crater II and Antlia II).  One halo\ (haloID=395, at z=0, hereafter halo395 ) of 231 MW like candidates, possesses a satellite plane as spatially thin and kinematically coherent as the observed one has been found. Halo395 resembles the MW in a number of intriguing ways: it hosts a spiral central galaxy  and its satellite plane  is almost ($\sim 87^{\circ}$)perpendicular to the central stellar disk. In addition, halo395 is embedded in a  sheet plane, with a void on the top and bottom, similar to the local environment of MW. More interestingly, we found that the 11 of 14 of the satellites on the plane of halo395, arise precisely  from the  peculiar geometry of its large-scale environment\ (\eg\  sheet and voids).  The remaining three members  appeared at the right place with the right velocity by chance at z=0. 
% Although the satellite plane of halo395 is transient and came into existence at z=0, the MW-like Large-scale environment indeed promotes the formation of the satellite plane.  
Our results support previous studies wherein the Satellite Plane Problem is not seen as a serious challenge to the $\Lambda$CDM model and its formation is ascribed to the peculiarities of our environment.   

\end{abstract}

% \keywords{Classical Novae (251) --- Ultraviolet astronomy(1736) --- History of astronomy(1868) --- Interdisciplinary astronomy(804)}

\section{Introduction} \label{sec:intro}

The accepted cosmological paradigm, known as the $\Lambda$CDM model, has been immensely successful at describing a myriad of observations. As a theory rooted in empirical observation and based on first principles, it can account for the formation of cosmic structures on a variety of scales: from galaxies like the Milky Way, to clusters and ultimately to the large scale structure of the Universe. Its success however has been challenged by observations of the smallest cosmological objects: dwarf galaxies. The careful observations of these objects in the vicinity of the Milky Way, when compared against theoretical models for how we expect them to form, have thrown up a number of problems or challenges \ (\eg\   reviews \citet{Bullock2017,Laura2022} ). The Satellite Plane Problem may be the most stubborn one\ \citep{Kroupa2005,Kroupa2012,Bullock2017,Pawlowski2018,Pawlowski2021Na,Boylan2021Na,Laura2022,Sawala2022Na,Kanehisa2023}. Although it was proposed decades ago\ \citep{Kunkel1976,Lynden1976}, its solutions are  still being debated. \par

The Satellite Plane Problem concerns the existence of a thin rotating plane of satellites, star clusters and streams, found around the MW\ \citep{Kunkel1976,Lynden1976,Kroupa2005,Metz2008,Pawlowski2013,Pawlowski2014,Pawlowski2015,Fritz2018,Pawlowski2020,Li2021ApJ} and other galaxies \ (\eg\  M31\ \citep{Conn2013,Ibata2013,Pawlowski2021apj}, Centaurus A\ \citep{Tully2015,Mueller2021,Kanehisa2023}). No matter what simulation is used\ (Collisionless, Hydrodynamic cosmological simulation\ \ (\eg\  \citet{Ahmed2017,Mueller2021,Pawlowski2020}) or zoom-in simulations which focus on the MW or Local Group\ (\eg\  \citet{Forero-Romero2018,Pawlowski2019,Santos2020,Samuel2021}) ) such a phase space configuration is always rare and transient\citep{Buck2016, Shao2019, Muller2021, Samuel2021,Sawala2022Na}.\par

Despite the fact that such configurations exist rarely, many mechanisms have been proposed to understand the formation of the thin rotating planes. For example,   \citet{Zentner2005,Libeskind2005,Shao2018} argue that satellites accreting along the  filaments will naturally lead to planar  configuration, while others \ (\eg\  \citet{Pawlowski2012}) disagree. \citet{Li2008mnras,Santos2020} suggest that accretion in a group may explain the kinematic coherence of the plane. Furthermore, recent studies\ \citep{Samuel2021} have shown that the presence of Large Magellanic Cloud(LMC)-like satellites will increase the chances of forming  planar structures, since many smaller satellites will fall together with LMC-like objects. Besides that, some studies\ \citep{Bilek2018,Bilek2021b,Banik2022} have proposed a more radical solution: satellite planes are generated from the tidal tails of interacting galaxies in the context of modified Newtonian dynamics instead of  $\Lambda$CDM.\par

 Previous studies on the small-scale challenges that use hydro-simulations, always have to make a choice: either a large cosmological box or zoom-in simulations. On one hand larger simulated volumes lack  sufficient resolution to study the dwarf galaxies at the heart of the small scale challenges, while on the other hand smaller volumes necessarily have smaller sample sizes thus limiting the scope of such studies. { It is also worth noting that the recent N-body simulations can guarantee both high resolution and large volume (\eg\  \cite{Ishiyama2021MNRAS}). However, we choose to use hydro-simulations in this work. Because they include baryonic processes, we can try to find a more realistic MW-like system: a thin rotating satellite plane around and perpendicular to the central spiral stellar disk. } The recently released hydro-simulation TNG50-1\citep{Nelson2019a,Nelson2019b,Pillepich2019}\ (Section \ref{sec:method})  balances resolution and sample size. The IllustrisTNG project is a suite of state-of-the-art cosmological hydro-dynamical simulations with well calibrated baryonic sub-grid physics implemented to reproduce the properties of local galaxy population, such as galaxy morphology and the stellar mass function\ \citep{Springel2018}.  
 % The  TNG50-1 is the largest simulation with sufficient resolution to address SP. 
 In this work, we aim to study the prevalence, origin and  time evolution of the MW's Satellite Plane by using TNG50-1.

 The structure of this article is organized as below: The observation data, numerical data and all kinds of measures we used in this work are enumerated in \Se{method}. \Se{select} presents the criteria used to select the model halo. Detailed  analyses of the selected model halos are described in \Se{analysis}. \Se{discuss} gives a  discussion. Finally, \Se{summary} summaries the main findings.  

\section{Methods} \label{sec:method}

\subsection{Observation data.} \label{sec:obs}

The 14 brightest MW satellites \ (11 classical satellites plus Canes Venatici I\ (CVn I), Crater II and Antlia II) are considered in this work. Although  many more satellites are observed around the Milky Way, the census at lower stellar mass is not observationally complete and the simulation can't resolve these masses in any case(see \Se{sim}). Many previous studies (\ie \ \citet{Koposov2007,Tollerud2008,Walsh2009,Carlsten2020,Carlsten2021} )have assumed the census of satellites with $\Mstar \ge 10^{5}\Msun$ is complete throughout the virial volume, while some disagree\citep{Yniguez2014,Samuel2020}. Here we follow the accepted assumption that the satellite sample with $\Mstar \ge 10^{5}\Msun$ or $M_{V}\lesssim -8$ is complete. Thus the brightest 14 satellites serve as the point of convergence between simulation resolution and observational completeness. \par

For the 11 classical satellites and  CVn I,  sky coordinates, heliocentric distance and line-of-sight heliocentric velocities are taken from \ \citet{McConnachie2012}, while for Crater II and Antlia II, these data are taken from\ \citet{Torrealba2016,Torrealba2019}. Proper motion data for all the 14 satellites are taken from\ \citet{McConnachie2020} which uses the Gaia Early Data Release-3 \ (EDR3). In order to estimate the observed uncertainties of some quantities\ (\eg\ ${\rm c/a,\  \Delta(k)}$. see the following section), the same methods as \ \citet{Samuel2021} are used here. For each satellite, the heliocentric distance, line-of-sight heliocentric velocities and proper motion are sampled 1000 times by assuming a Gaussian error distribution, then converted to a 6-D \ (positions and velocities) Cartesian Galactocentric  coordinates by using Astropy\ \citep{Astropy2013,Astropy2018}.\par

% For the TBTF problem, as in previous study\ \citep{Boylan2012}, we only consider the 9 MW satellites: 8 classical satellites (except Sagittarius, LMC, SMC) plus  CVnI. Because Sagittarius is interacting strongly with the MW stellar disk, then the circular velocity implied from the velocity dispersion may be not accurate, so it was
% removed. As for the SMC and LMC, The model halo does not own SMC/LMC analogues, so they were removed too. In order to get the circular velocity at the 3-D half-light radius ${\rm V_{circ}(r_{1/2})}$, the  formula in\ \citet{Wolf2010} was used to calculate ${\rm V_{circ}(r_{1/2})}$. The 3-D half light radius and stellar velocity dispersion of each satellite were taken from  \ \citet{Wolf2010}.\par

\subsection{Numerical Simulation.} \label{sec:sim}

Throughout this work, we compare observations with the simulation data from TNG50-1. This has a box size $\sim 51.7\Mpc$,  dark matter resolution $\sim 4.5\times 10^{5}\Msun$ and stellar particle resolution $\sim 5\times 10^{4}\Msun$. It adopts the best fitting cosmological parameters in \ \citet{Planck2016}: dark energy density $\Omega_{\Lambda} = 0.6911$, matter density ${\rm \Omega_{m} = 0.3089}$, baryon density ${\rm \Omega_{b} = 0.0486}$, Hubble constant ${\rm H_{0} = 67.74kms^{-1}Mpc^{-1}}$, normalization ${\rm \sigma_{8} = 0.8159}$ and the spectral index ${\rm n_{s} = 0.9667}$. The compromise between sample size and simulation resolution of TNG50-1 make it ideal for analyzing the satellite galaxies in MW-mass halos. 
 
\subsection{Measures of disc thickness.}\label{sec:thickness}
We use two methods to measure the thickness of the satellite plane:  minor to major axis ratio\ (${\rm c/a}$) and root-mean-square\ (RMS) height ${h}$.\par 

Ratio ${\rm c/a}$ is calculated as the ratio of square root of the minimal and maximum eigenvalues of the inertia matrix M:
\begin{equation}
    {\rm
    \boldsymbol{M} = \sum_{i} (\boldsymbol{r}_{i}-\boldsymbol{r}_{0})^{T} (\boldsymbol{r}_{i}-\boldsymbol{r}_{0})
    }
\end{equation}

where ${\boldsymbol{r}_{i}}$ is the position vector of ${\rm i^{th}}$ satellite relative to the halo center and $\boldsymbol{r}_{0}$ is the geometric center of  the system of  satellites: ${\rm \boldsymbol{r}_{0}=\frac{1}{N_{sat}}\sum_{i}\boldsymbol{r}_{i}}$. The normal of the plane ${\rm \hat{\boldsymbol{n}}_{c}}$ is defined as the direction of the minor axis \ (\ie\ the eigenvector corresponding to the smallest eigenvalue c). \par

As for the RMS height ${\rm h}$, we use a similar definition as \ \citet{Samuel2021}. We randomly generate $10^{4}$ planes centered on the geometric center of satellites $\boldsymbol{r}_{0}$, then iteratively calculate the RMS height:
\begin{equation}
    {
    h = \sqrt{\frac{\sum_{i=1}^{N_{sat}}(\hat{n}_{\bot}\cdot\vec{x}_{i})^{2}}{N_{sat}}}}
\end{equation}
where $\hat{n}_{\bot}$ is the unit normal vector of the random plane and $\vec{x}_{i}$ is the position vector of ${\rm i^{th}}$ satellite with the origin is on the geometric center of those satellites. The minimum value of h is the RMS height of satellites and the corresponding plane is defined as the midplane. The smaller the values of ${\rm h}$ and ${\rm c/a}$, the thinner the plane.

\subsection{Measures of kinematic coherence.}
We use the ${\rm k-\Delta_{k}}$ relation\ \citep{Pawlowski2020} to characterize the kinematic coherence. For a given group of ${\rm N_{sat}}$ satellites, we flip the direction of angular momentum of retrograde satellites at first\ (\ie\ $(-1) \cdot \boldsymbol{\rm n_{retrograde}}$), as we treat retrograde and prograde satellites equally here. Specifically, we first randomly generate 1000 direction vectors. For each random direction ${\rm \boldsymbol{n}_{ random,\ j},\ (j= 1,2,\cdots,1000})$ , we calculate the RMS angle distance $\rm D_{j}$ between the angular momentum direction of those ${\rm N_{sat}}$ satellites and the random direction ${\rm \boldsymbol{n}_{ random,\ j}}$  based on the formula: 
    \begin{equation}
        {\rm D_{j}} = \sqrt{\frac{\sum^{\rm N_{ sat}}_{i=1}[\arccos{(|\boldsymbol{n}_{\rm random,\ j}\cdot\boldsymbol{n_{i}}|)}]^2}{\rm N_{sat}}}
        \label{eq:eq2}
    \end{equation}  
    where ${\boldsymbol{n_{i}}}$ represents the angular momentum direction of ${\rm i^{th}}$ satellite and we note that there is a absolute sign in the formula. Then the random direction with the  minimal angle distance \ (\ie\ $ \min {\rm \{D_{j}\}}$) is defined as the best direction ${ \boldsymbol{\rm n_{best}}}$  and  the angular momentum of a satellite would be flipped if the angle between the direction of its angular momentum and the best direction {$ \boldsymbol{\rm n_{best}}$}\ (\ie\  $  \arccos{(\boldsymbol{\rm n_{best}} \cdot \boldsymbol{\rm n_{i}})}$)  is larger than $90^{\circ}$. After flipping the angular momentum of retrograde satellites, the ${\rm k-\Delta_{k}}$ relation is obtained through the following steps:

\begin{enumerate}
    \item For a positive integer ${\rm k}$ with ${\rm k\leq N_{sat}}$, we can get ${\rm \binom{N_{sat}}{k}}$ different combinations of k satellites among the $\rm N_{sat}$ samples.
    \item We calculate ${\rm \Delta(k)}$ for each combination according to the following formula:
    \begin{equation}
        \Delta(k) = \sqrt{\frac{\sum^{k}_{i=1}[\arccos{(\langle\boldsymbol{n}\rangle_{k}\cdot\boldsymbol{n_{i}})}]^2}{k}}
        \label{eq:eq1}
    \end{equation}
    Where $\boldsymbol{n_{i}}$ is the angular momentum direction of $i^{th}$ satellite and ${\rm \langle\boldsymbol{n}\rangle_{k}}$ represents the average direction of the ${\rm k}$ satellites.
    \item Then ${\rm \Delta_{k}}$  is defined as the minimal ${\rm \Delta(k)}$  \ (\ie\ ${\rm \Delta_{k} \equiv \min \{\Delta(k)\}}$).
\end{enumerate}
Repeating those processes for a series of $\rm k$ values, we will get curve ${\rm k-\Delta_{k}}$. Based on this curve, only those model galaxies which meet ${\rm \Delta_{k,model}\leq {\rm 84th \ percentiles\  of\ }\Delta_{k,obs} (\ie\ \leq median + 1\sigma ),}$\  ${\rm \forall k \in [4,5,\cdots,14]}$  are considered to be kinematically comparable to MW.\par

In addition to the angular dispersion ${\rm \Delta_{k}}$, we also identify whether the satellites are orbiting in the plane they formed. This is important as it determines if this planar configuration is stable for a long time. We calculate angles ${\rm \theta_{k}}$ between the normal of the plane ${\rm \hat{\boldsymbol{n}}_{c}}$ and the average angular momentum  direction ${\rm \langle\boldsymbol{n}\rangle_{k}}$ of the ${\rm k}$ satellites which have minima $\Delta(k)$ among all the ${\rm \binom{N_{sat}}{k}}$ combinations \ (\ie\ ${\rm \min \{\Delta(k)\}}$):

\begin{equation}
    {\rm 
    \theta_{k} = \arccos{(|\hat{\boldsymbol{n}}_{c}\cdot \langle\boldsymbol{n}\rangle_{k}|)},\ k \in [4,5,\cdots,14]
    }
\end{equation}

% \subsection{Luminosity function.}
% When calculating the satellite luminosity function, all luminous satellites with ${\rm subhaloFlag=1}$(which determines whether a subhalo is likely a cosmological origin or from numerical effects) and  within $300\Kpc$ were included. Then the luminosity function of the model galaxy (i.e., Halo 395) is compared with the recent result on the luminosity function of the MW\ \citep{Newton2018}, as shown in the first panel of \Fg{fig1}.

% \subsection{Rotation curve.}

% When plotting the rotation curve, we selected all subhalos with $\Mstar \ge 10^{5}\Msun$, within $300\Kpc$ and with ${\rm subhaloFlag=1}$. In order to compute the rotation curve for each subhalo with minimal numerical effects, we follow the work in\ \citet{Boylan2012}. Firstly, we fit the mass density profile of a subhalo to its well-resolved radii, then obtain the rotation curve based on the fitted density profile. In our work, the inner radius of the well-resolved part was determined by the criteria proposed by the study\ \citep{Power2003}. However, for some subhalos with lower number of particles (DM $+$ star: $\sim 80-1600$), the fit is unreliable. For these subhalos, we directly compute the rotation curve using the simulated raw particle data based on the formula:
% \[\rm V_{circ}(r)= \sqrt{\frac{GM(<r)}{r}}\]

\section{Selecting model halos} \label{sec:select}

The first step in attempting to find systems to compare with the observations is to identify MW ``like'' halos in the simulation at $z=0$. These are \ (generously) selected with 
$ {\rm M_{200}\in [0.3,3]} \times 10^{12}\Msun$ and  ${\rm \Mstar\in [1,10]\times10^{10}\Msun}$, where  $M_{200}$ is the total mass of the halo enclosed in a sphere with an average density 200 times the critical density of the universe and $\Mstar$ is the  total stellar mass of the central galaxy  within twice the stellar half mass radius. An additional criterion is applied to ensure that the MW halos that have been identified are located in environments similar to ours. An isolation requirement is imposed that ensured no halo more massive than $0.5\times 10^{12}\Msun$ exists within $780\Kpc$, the distance between the MW and M31. Considering both the observational completeness limits of the census of the satellite galaxy population and the resolution of TNG50-1, we select the top 14 brightest satellites \ (as done by \ \citep{Samuel2021}), which have $\Mstar \ge 10^{5}\Msun$ and are between  $15\Kpc {\rm\ and\ } 300 \Kpc$ from the center of the host galaxy.  Any halo which doesn't meet these criteria (\ie doesn't have at least 14 satellites in this volume) is omitted from consideration.  For the remaining halos, the 14 satellites with the most massive stellar mass are selected for further analysis. For reference: 548 halos in the simulation fall into  the stated halo and stellar mass range; of these 404 are isolated and have no massive halo closer than the distance to Andromeda; of these 231 have at least 14 satellites that are above the stated stellar mass limit, and sit outside the host galaxy disc but within 300kpc.

{\bf Only one single halo} at ${\rm z=0}$  survives all these constraints and matches the thinness and orbital coherence of the MW system. Namely it is as massive as the MW in both DM and stellar mass, is isolated according to the criteria stated earlier and has a similarly thin and co-rotating satellite structure. The ID of the halo is 395 \ (\ie\ haloID=395, at ${\rm z=0}$).  In the following we present detailed analysis of this halo395 and investigate how its satellite plane forms in the simulation.

\section{Analyzing Halo395}\label{sec:analysis}

\subsection{Basic Characteristics} \label{sec:basicChar}

Halo395 has a dark matter mass ${\rm M_{200}=4.9\times 10^{11}\Msun}$, which is at the lower limit of observational constraints on the MW halo mass\ \citep{Wang2020}, though some recent results favor a lower MW halo mass\ (\eg\ \cite{Bird2022MNRAS}). Its stellar mass is $\Mstar = 2\times 10^{10}\Msun$, which is similar to the accepted values for the MW \ (e.g. $\Mstar = 6\times 10^{10}\Msun$\ \citep{Licquia2015}). The halo hosts a spiral disk galaxy as well\ (see panel(a) of \Fg{fig2}). The stellar mass \ (within two half-light radii) of the most luminous satellite is $\sim 3\times10^{7}\Msun$, thus this model galaxy doesn't possess satellites as massive as the Large Magellanic Cloud \ (LMC), whose mass is estimated at $\Mstar\sim 1.1\times 10^{9} \Msun$ or the Small Magellanic Cloud\ (SMC), whose mass is estimated at $\Mstar\sim 3.7\times 10^{8} \Msun$ \ ( see table A1 of \citet{Garrison-Kimmel2019}). %and of the faintest satellite is $\sim 2.6\times10^{5}\Msun$, made of just 5 star particles. 
% {\color{red} As the only model galaxy that has a SP similar to that of the MW, the other two issues (TBTF and MS) were also checked. Surprisingly, these problems appear to be resolved as well. Below, all three of these issues are presented}.\par

%%%%%%%%%%%%%%%%
%% figure

\begin{figure*}[ht!]
    \includegraphics[scale=0.5]{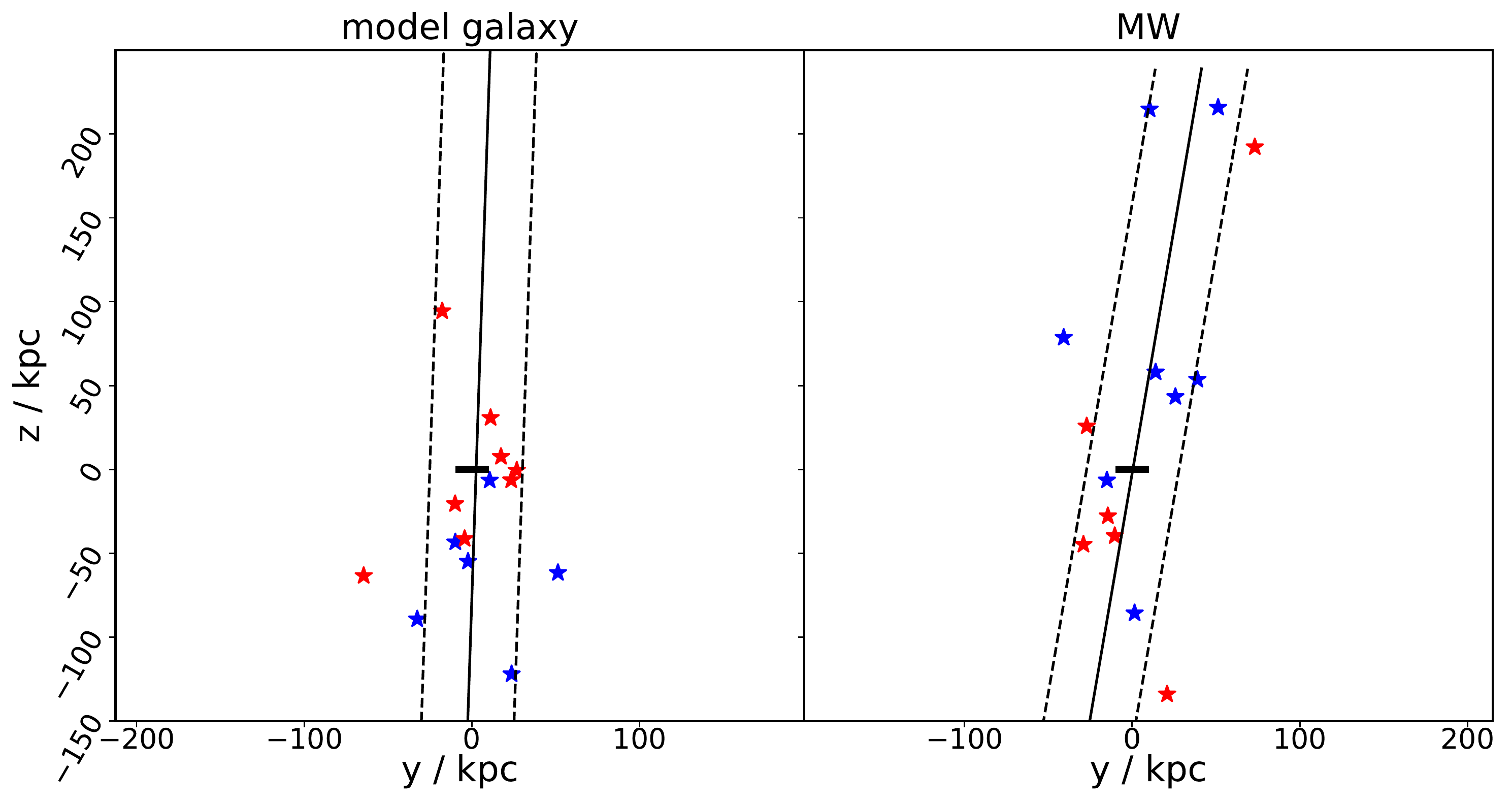} 
    \centering
    \caption{\textbf{Edge-on view of the satellite plane of halo395 and MW.}
     This figure is for Satellite Plane problem, which shows the positions of the top 14 brightest model/MW satellites\ (blue/red stars), and central stellar disk\ (thick black solid line). The view is oriented to show the edge-on of both the central stellar disk and the satellite plane \ (long black solid line). Color blue/red indicate that satellites are moving out of/into the screen. The long black dashed lines indicate the root-mean-square height of the plane \ (Section \ref{sec:method}). The left/right panel depict  halo395/MW.  
    } 
    \label{fig:fig1}
\end{figure*}

%%%%%%%%%%%%%%%%%%%%%%%%%%

% \subsection{MS and TBTF Problem}
% \Fg{fig1} demonstrates that halo 395 does not suffer from the three small-scale problems that are known to haunt $\Lambda$CDM. The left panel shows that there is no MS problem in this galaxy, as the model luminosity function matches well the observed satellites of the MW (except at the luminous end where the model galaxy has no analogue of the SMC and LMC). Although the resolution is insufficient to probe the mass function below $M_{\rm h200}\sim10^{6}M_{\odot}$ or $M_{\rm star}\sim10^{5}M_{\odot}$ - the number of subhaloes and satellites with mass greater than this is in good agreement with the observations. The right panel of this figure  shows that there is also no TBTF problem for this galaxy either. The lines are the rotation curves of massive satellite galaxies (those with $\Mstar\ge 10^{5}\Msun$). The circular velocity of observed satellites are in agreement with the rotation curves of the simulated satellites, and there are no massive failures in the model.
%It has been shown that it is abnormal for the MW to host such luminous satellites\cite{Liu2011}. 

%%%%%%%%%%%%%%%%%%%%
%% figure
\begin{figure*}[ht!]
    % \plotone{/fig4.pdf} 
    \includegraphics[scale=0.9]{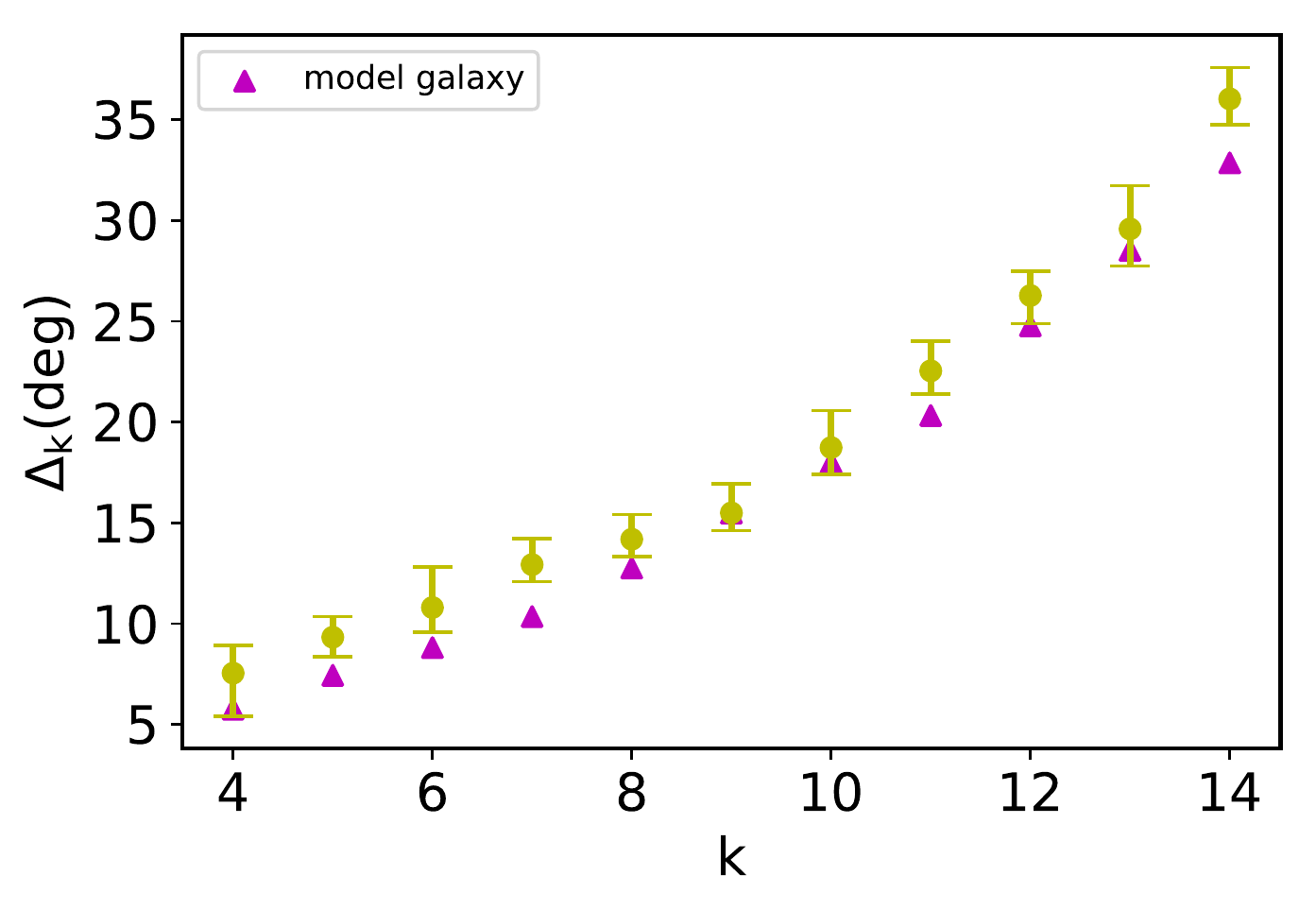} 
    \centering
    \caption{\textbf{the ${\rm \bf k-\Delta_{k}}$ curve.}
    Yellow data points with error bars show the observational results of the MW with $1\sigma$  uncertainty. The magenta triangles are for the model galaxy, halo395.
    }
    \label{fig:fig4}
\end{figure*} 

\begin{figure*}[ht!]
    % \plotone{/fig5.pdf} 
    \includegraphics[scale=0.9]{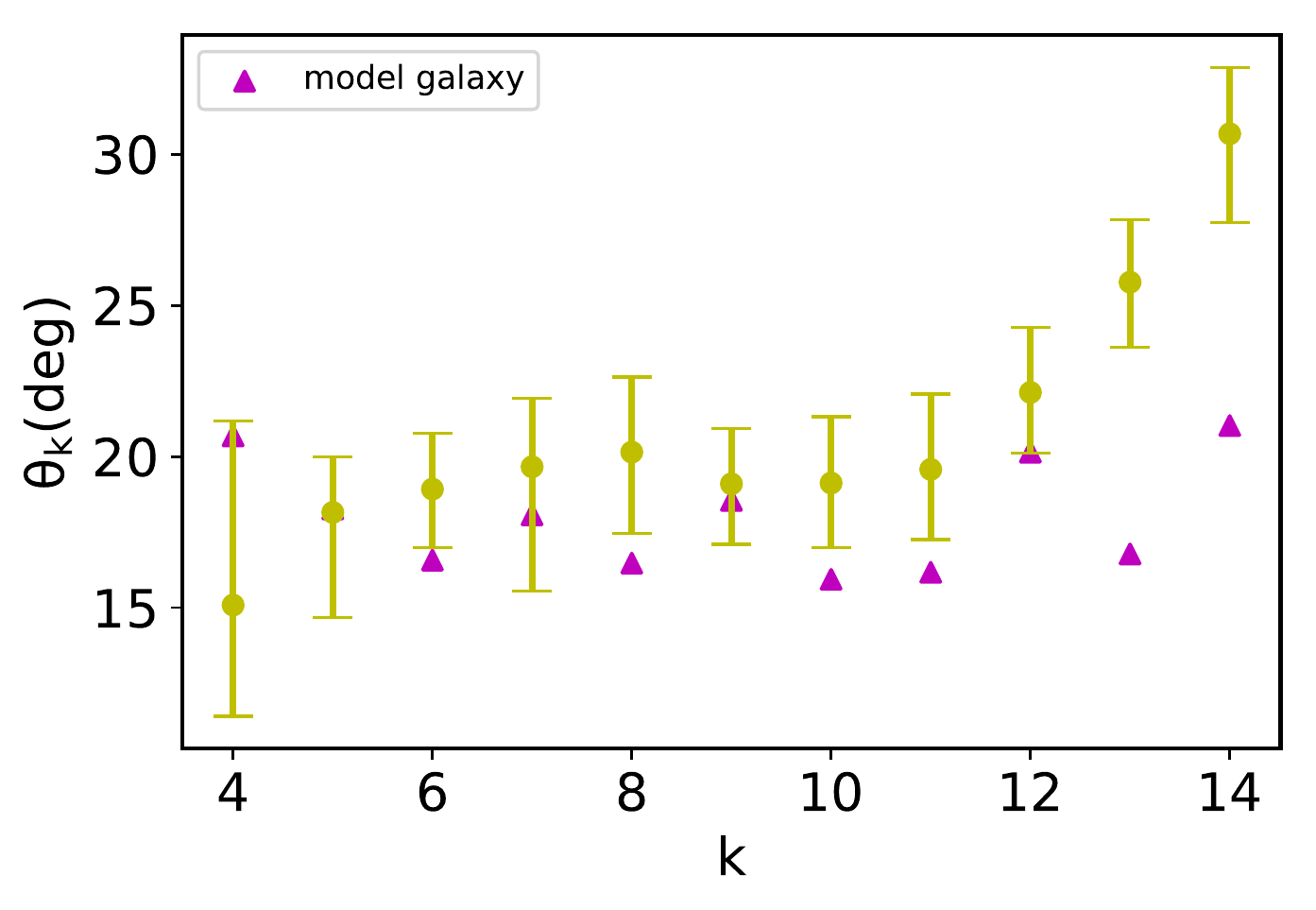} 
    \centering
    \caption{\textbf{the ${\rm \bf k-\theta_{k}\ relation}$.}\ 
    Yellow data points with error bar show the observational result of the MW with $1\sigma$  uncertainty, and magenta triangles for the model galaxy. Here $\theta_{k}$ is the angle between the average of the satellite angular momentum and the normal of the satellite plane. This quantity measures how well is the orbital plane aligned with the spatial plane. 
    }
    \label{fig:fig5}
\end{figure*}

\begin{figure}[ht!]
    \includegraphics[scale=0.7]{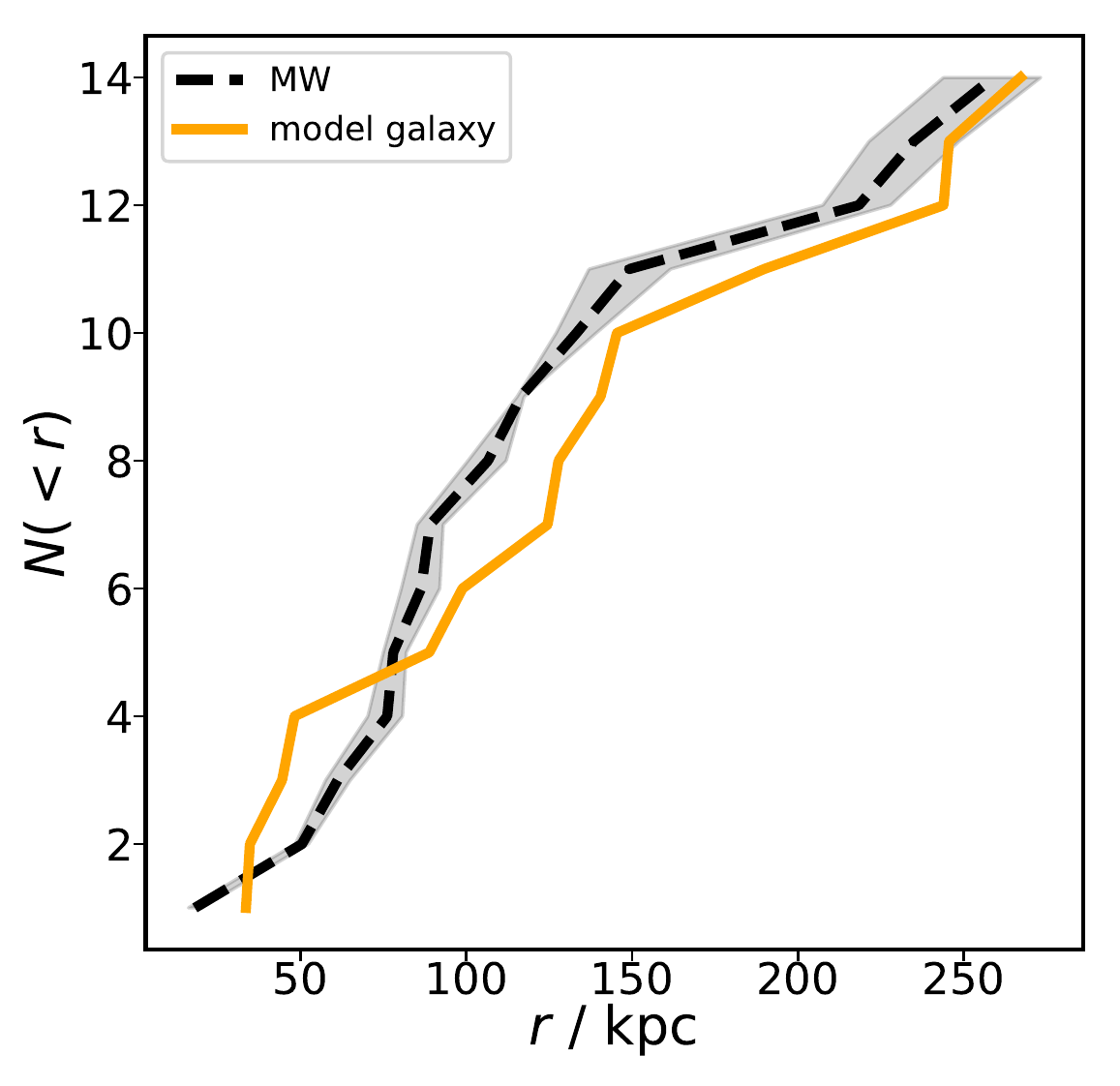}
    \centering
    \caption{\textbf{the satellites radial distribution of MW and halo395. }
    The black dashed line with gray shaded region shows the observational result of MW with $1\sigma$ uncertainty, and the orange solid line is that of halo395. Apparently, the satellites radial distribution of halo395 broadly match the observation.
    }
    \label{fig:fig8}
\end{figure}

%%%%%%%%%%%%%%%%%%%%

\subsection{The Satellite Plane Problem }
 \Fg{fig1} shows an edge-on view of the satellite plane (both halo395 and MW), with satellites colored red or blue depending on whether their tangential velocity is approaching or receding in this projection.  From this plot the satellites of halo395 under consideration are  distributed in a thin plane, and most satellites are orbiting in this plane in a manner that is consistent with rotation. The ratio $c/a$  and RMS height $h$ of this halo's satellite population is 0.2 and $27\Kpc$ respectively while, in the case of observed MW, the values are 0.25 and $27\Kpc$ respectively . As for orbital coherence, the curve ${\rm k-\Delta_{k}}$ of the model halo is below that of observed one when considering a $1\sigma$ uncertainty\ (see \Fg{fig4} and Section \ref{sec:method}).  \Fg{fig5} shows the $\rm k-\theta_{k}$ relation of the model galaxy and of observation with $1\sigma$ uncertainty. Obviously, the model galaxy behaves better than the MW at all values of ${\rm k}$. Furthermore, the angle between the average orbit pole of the satellites and the norm of the satellite plane is also $\sim 21^{\circ}$ which is smaller than the observed value of $\sim 31^{\circ} $ \ (see \Fg{fig5}). Besides that,  in \Fg{fig8}, the satellites radial distributions of halo395 and MW are compared. It is clear that the satellites radial distribution of halo395 is roughly consistent with that of the MW.

 The satellite plane of halo395 is almost perpendicular \ ($\sim 87^{\circ}$)  to the central stellar disk \ (see the left panel of \Fg{fig1}). In the case of the observed MW, the angle between the satellite plane and the MW disk is $\sim 81^{\circ}$(see the right panel of \Fg{fig1}). Thus not only does this halo have a thin rotating plane, but it is oriented in the same sense as the observations namely roughly perpendicular the MW's disk.  In sum: halo395 which resembles the MW in terms of its halo mass, its stellar mass, its morphology, its isolation and its satellites radial distribution, also bears a thin satellite plane that's spinning in a coherent manner  and is oriented exactly in the same way with respect to the Galaxy itself.

%%%%%%%%%%%%%%%%%%%%%%%%%%
%% table

% \begin{table*}
% \centering
% \caption{}
% \label{tab:Apex}
% \begin{tabular}{ccc}
% \hline\hline
% \multicolumn{1}{c}{} & Observed & Simulated \\
% Mass of MW & $[0.5,2]\times10^{12}\Msun$\citep{Wang2020} & $0.49\times10^{12}\Msun$\\
% Mass of closest neighbour (M31) & $[0.6,2]\times10^{12}\Msun$\citep{Diaz2014,Corbelli2010,Tamm2012,Kafle2018} & $0.52\times 10^{12}\Msun$\\
% Distance to closest Neighbour (M31) & $770\pm40\ \mathrm{kpc}$\citep{Holland1998,Joshi2003,Ribas2005,McConnachie2006} & $790\ \mathrm{kpc}$\\
% Mass of 2nd closest neighbour (CenA) & $1.2\times10^{13}\Msun$\citep{Muller2022} & $1.4\times 10^{13}\Msun$\\
% Distance to 2nd  closest Neighbour (CenA) & $3.8\pm0.1\ \mathrm{Mpc}$\citep{Harris2010} & $2.6\ \mathrm{Mpc}$\\
% \end{tabular}
% \end{table*}

%%%%%%%%%%%%%%%%%%%%%%%%%%%%%%%%%%%%%%%%%%%

% As there are no lines(rotation curve of massive satellites) in the panel that are systematically higher than the observed points(\ie\ don't have strong massive failures) and only four curves(represent four subhalos) are as dense as Draco and Ursa Minor(the top 2 densest satellites of the MW), which all can be assigned a MW satellite(\eg\ Draco, Ursa Minor, Sculptor and LeoI), so the model halo also don't have massive failures\cite{Garrison2014}.}\par

%%%%%%%%%%%%%%%%%%%%%%%%%%%
%%%figure
\begin{figure*}[ht!]
    \includegraphics[scale=0.5]{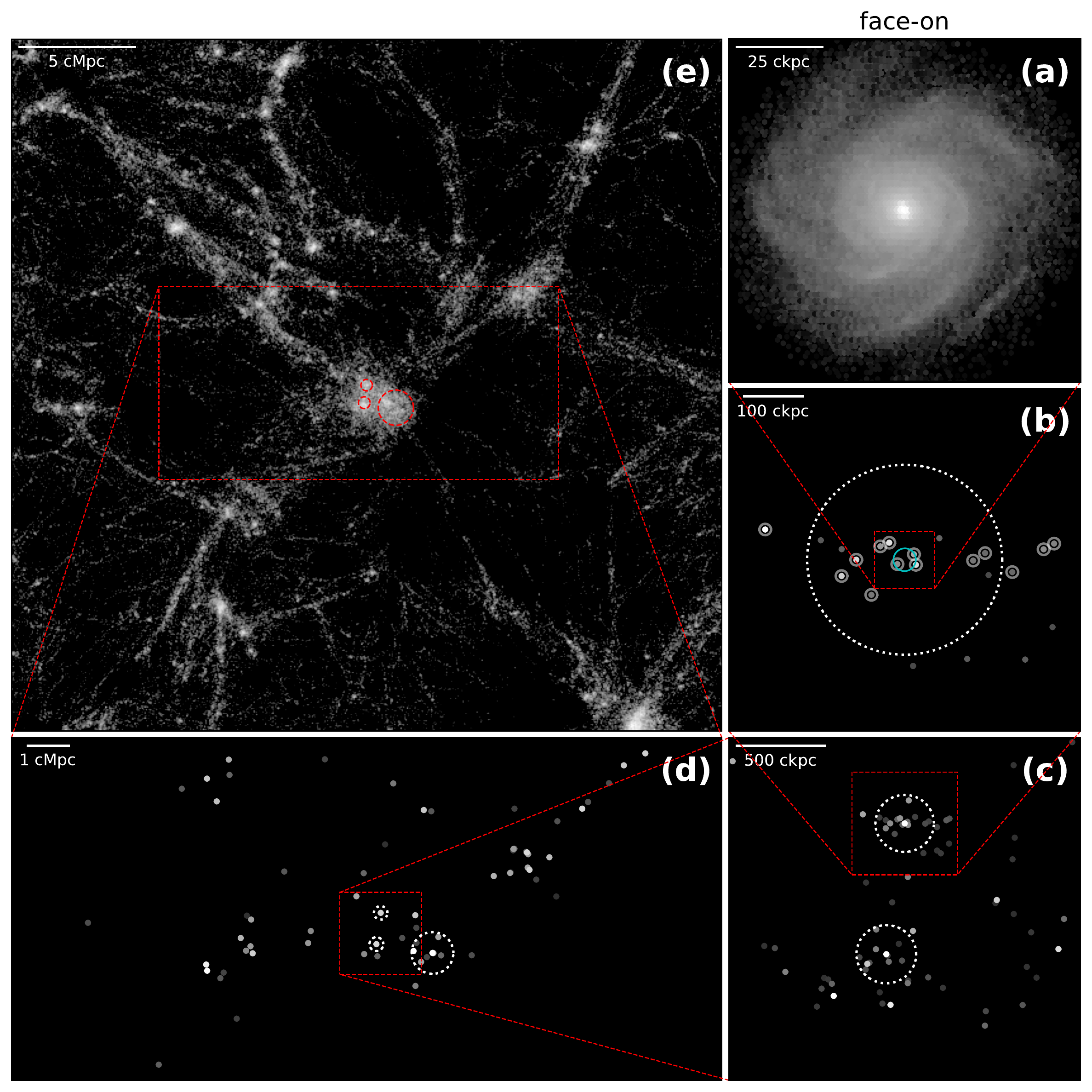}
    \centering
    \caption{\textbf{the zoom-in view of the model galaxy and its surroundings at z=0.}
    Here, the views of all panels are oriented to show the edge-on view of the satellite plane. From panel (a) to panel (e), the upper right panel shows the image of the central spiral galaxy of the model halo \ (halo395). Panel (b) shows the distribution of all luminous satellites \ ($\Mstar > 10^{5}\Msun$), and where the top 14 brightest satellites are marked by white solid circles. The central cyan solid circle indicates the central stellar disk, which is almost perpendicular to the satellite plane. Panel (c) shows a `M31' analogue \ (${\rm M_{200} = 5.2\times 10^{11}\Msun}$). The `Centaurus A/M83 group' \ ( with halo mass ${\rm M_{200} = 1.4\times 10^{13}\Msun}$ ) is shown as big circle in  panel (d) where the points represents all subhalos with $\Mstar > 10^{8.5}\Msun$. Panel (e) shows the large-scale environment around the model galaxy. The dotted circles in each panel are scaled by the virial radius of the halos in that panel. The whiteness of points in each panel are scaled to the stellar mass of the satellites. The thickness of slice (e) and (d) are 3 ${\rm cMpc}$ and 8 ${\rm cMpc}$ 
    % (as only very luminous subhalos were included in panel(b))
    respectively, where ${\rm cMpc}$ is the co-moving radius. 
    } 
    \label{fig:fig2}
\end{figure*}

%%%%%%%%%%%%%%%%%%%%%%%%%%%

\subsection{Environment of the Halo}
More surprising results are seen in \Fg{fig2} which shows the environment on various scales of the halo in question. Panel (a)  shows that halo395 hosts a spiral stellar disk very similar to our MW disk. Panel (b) shows the satellites of the halo. It clearly displays that most luminous satellites form a plane which is perpendicular to the central stellar disk\ (the green circle, face-on view). Even more surprising is that there are `M31' and `Centaurus A/M83' analogues living in the neighborhood of the model galaxy, which are shown in panel (c) and (d). The `M31' and `Centaurus A/M83' group are the only two objects in the neighborhood \ ($<4\Mpc$) which are more massive than halo395. Surprisingly, the distance to the `M31' \ (which has a halo mass of $\sim 1.06$ times the halo mass of halo395) and `Centaurus A/M83'\ (with ${\rm M_{200} = 1.4\times 10^{13} \Msun}$ while the observation is ${\rm M_{200} = 1.2\times 10^{13} \Msun}$\ \citep{Muller2022} ) are 790 $\Kpc$ and 2.6 $\Mpc$ respectively, which are very close to the data \ ($770\pm40\ \mathrm{kpc}$\ \citep{Holland1998,Joshi2003,Ribas2005,McConnachie2006} and $3.8\pm0.1\ \mathrm{Mpc}$\ \citep{Harris2010} respectively). Again, we note that our model `M31' also has a slightly lower mass than the real one \ ($\sim 5\times 10^{11}M_{\odot}$ vs $\sim [6-20] \times 10^{11}M_{\odot}$) but the mass ratio between the simulated and observed LGs is nearly identical. On large scales, this `local group' is embedded in a sheet with a void above and below, as shown in panel (e). 
%{\color{red} There is also a Virgo mass cluster ($\sim 10^{14}M_{\odot}$ located at XXX Mpc - albeit this is further than the 17Mpc distance to the observed Virgo. Still though it is ``in the neighbourhood'' and much closer than the mean cluster distance in $\Lambda$CDM}. 
In short, the model galaxy and its environment indeed resembles the cosmography of the MW and the local group\ \citep{Tully2019}.  \par

%%%%%%%%%%%%%%%%%%%%%%%%%%%
%%%figure
\begin{figure*}[ht!]
    \includegraphics[scale=0.6]{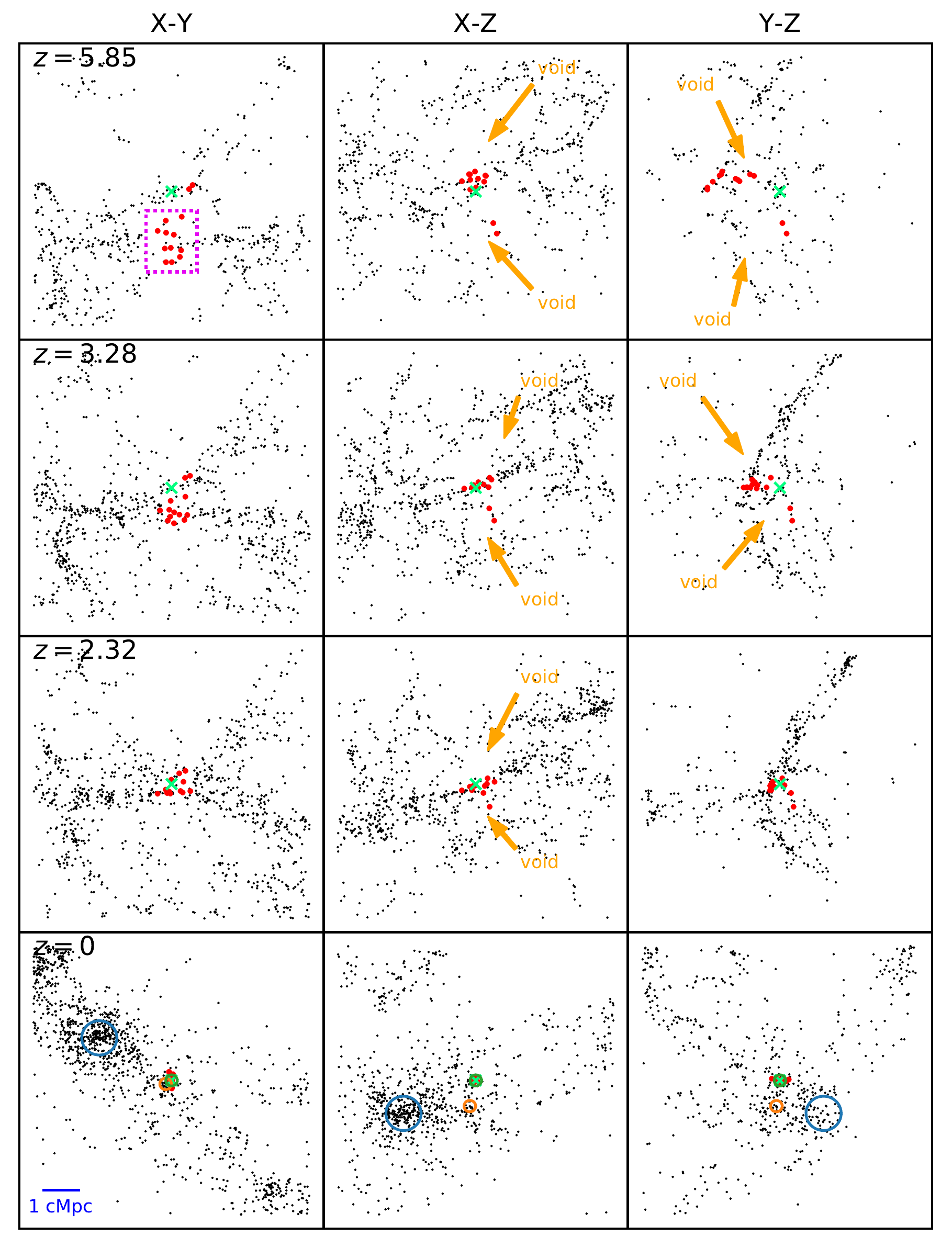} 
    \centering
    \caption{\textbf{the large-scale structure around the model galaxy and its evolution.}
    The view of this figure is rotated so that the normal of the satellite plane is aligned with the Z-axis. Each column shows the projection to the X-Y, X-Z and Y-Z planes. Each row is for different epoch \ (redshift: 5.85, 3.28, 2.32, 0 or Cosmic Age: 0.96, 1.94, 2.84, 13.8 Gyr) of the evolution.
    The green cross and the red dots represent the central galaxy and satellites \ (top 14 brightest) respectively. In all panels, only subhalos with total mass $\ge 4.5\times 10^{8} \Msun$ are shown. The blue, orange and green circle indicate the virial radii of `Centaurus A/M83 group', `M31' and the model galaxy. Red dots surrounded by a dotted rectangle in the upper left panel fell on the model galaxy along a sheet plane and are analyzed in \Fg{fig7}. It is worth noting that only 13 red dots were shown on the first three rows, as one of the 14 satellites formed very recently until $z=1.41$ or ${\rm Cosmic\ Age=4.5\ Gyr}$. The thickness of all slices \ (panels)  are 3 ${\rm cMpc}$.
    } 
    \label{fig:fig3}
\end{figure*}

\begin{figure*}[ht!]
    \plotone{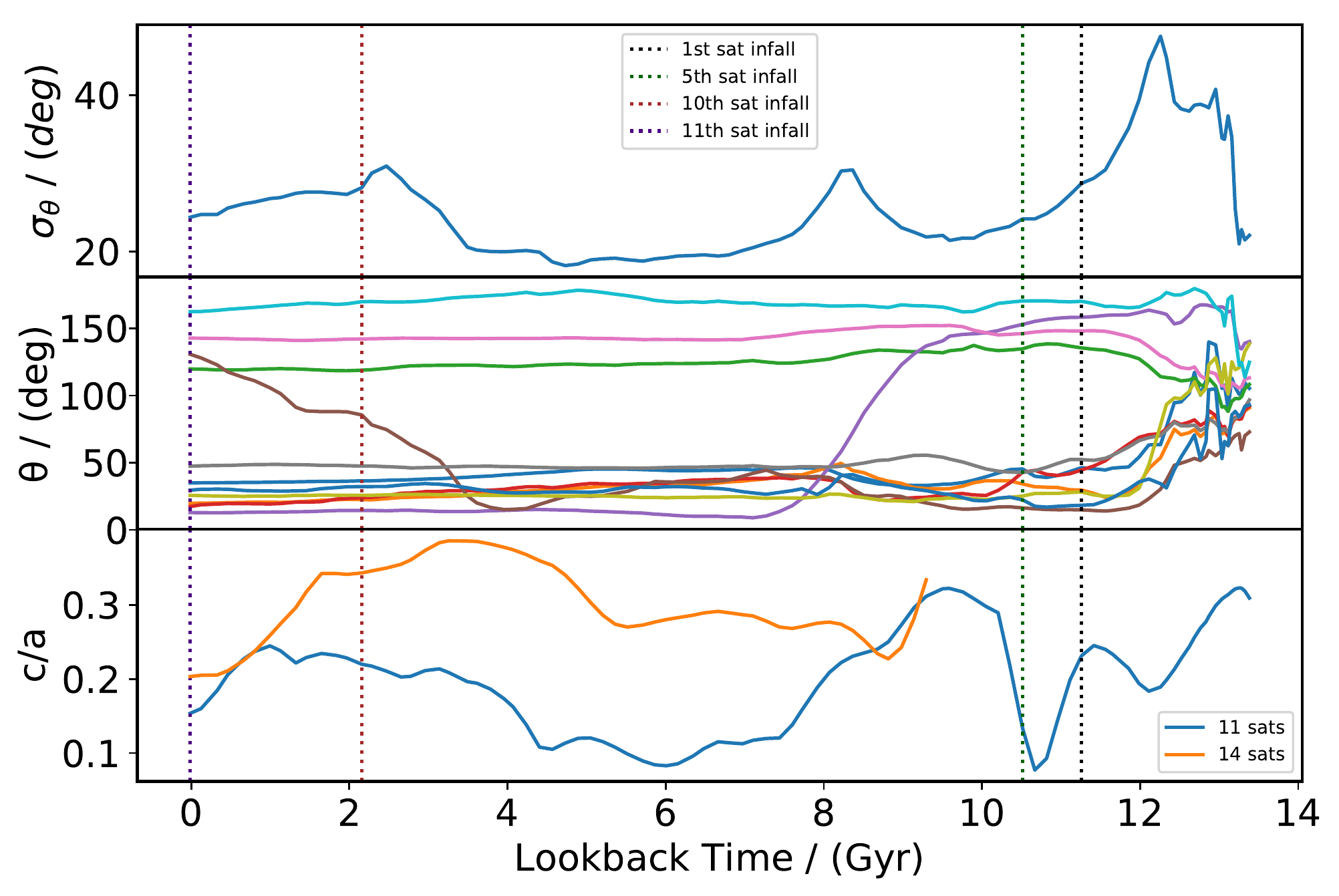} 
    \caption{\textbf{The evolution of  ${\rm \bf c/a}$, the  angular momentum of 11 satellites and their dispersion.}
    We consider 11 satellites which are the red dots surrounded by a dotted box in \Fg{fig3}, and we obtain the evolution of their individual angular momentum, $\theta$, and the dispersion of their angular momentum \ ($\sigma_{\theta}$). The upper panel shows the evolution of $\sigma_{\theta}$, where prograde and retrograde satellites are treated equally. The middle panel shows the evolution of angular momentum $\theta$ for each individual satellite. Here $\theta$ is measured between the angular momentum of a satellite and an arbitrary fixed direction vector. The lower panel shows the evolution of ${\rm c/a}$ of the satellite plane composed of 11(blue line) and full 14(orange line) satellites. The orange line starts at Lookback Time =9.5 Gyr, because one of the 14 satellites didn't form until then, see the caption of Figur\ \ref{fig:fig3}. Note that only here we consider all 14 satellites.  The four vertical dotted lines \ (with black, green, red and purple color) indicate the epochs when the 1st, 5th, 10th and the 11th \ (the last) satellite are accreted by the model galaxy respectively. The accretion time is defined as the time when the satellite first crossed the halo virial  radius. Actually, the 11th satellite never crossed the virial radius, but it was within $300 \Kpc$ at ${\rm z=0}$.  It is also worth noting that, in the middle  panel, the satellites represented by the three solid lines \ (with $\theta \sim 150^{\circ} $) at the top are retrograde in the plane, and others are prograde. It is seen that for most of the lookback time, the angular momentum of satellites are almost kept fixed since their accretion.
    % As for the sharp rise of the dispersion (LookBack Time ${\rm >12\ Gyr}$) in the top panel, it may be resulted from the strongly interaction between these satellites and the large scale structure at early time.
    }
    \label{fig:fig7}
\end{figure*} 

\begin{figure*}[ht!]
    % \plotone{/fig6.pdf} 
    \includegraphics[scale=0.9]{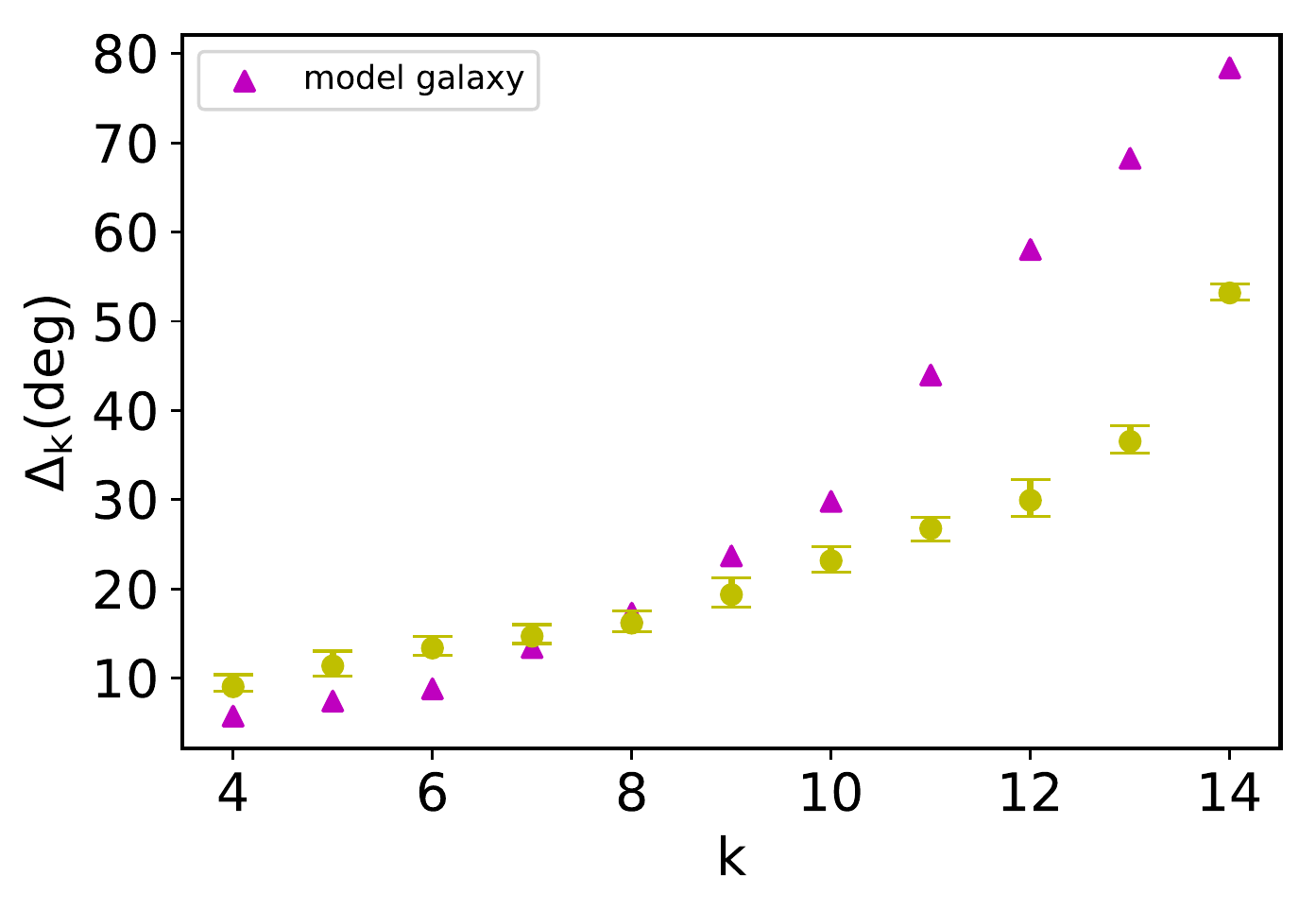} 
    \centering
    \caption{\textbf{the ${\rm \bf k-\Delta_{k}}$ curve, computed using \eq{eq1} without flipping the angular momentum of retrograde satellites.}
    All data points are similar to \Fg{fig4}.
    }
    \label{fig:fig6}
\end{figure*}

%%%%%%%%%%%%%%%%%%%%%%%%%%%
\subsection{Possible Origin of the Satellite Plane}\label{sec:origin}

Previous studies\ \citep{Garrison2019,Brooks2014,Dutton2016,Buck2019,Laura2022} have shown that other small-scale challenges can be  solved with well-designed subgrid models, while the Satellite Plane Problem has often been an obstacle and solutions are in debate. 
It was suggested\ \citep{Libeskind2015} that the plane of satellites originated from the geometry of the cosmic web and the same result was also obtained here. 
% Here, we show the evolution of the large-scale environment around the model galaxy in \Fg{fig3}. 
The upper panel of \Fg{fig3} shows that 11 of 14 satellites \ (marked with a dotted box) fell into the model galaxy along the plane of a large cosmic sheet. The dispersion of the satellite's angular momentum decreased as the sheet, they were formed in, become thinner due to the gravitational collapse of the sheet (and expansion of the voids above and beneath the sheet, see  \Fg{fig7} and \Fg{fig3}).  After most of (10/11) the 11 satellites enter  and evolve in halo395, their  angular momentum remain unchanged, as the potential of the halo is almost isotropic \ (see the middle panel of \Fg{fig7}), which leads to two important results: (1) The dispersion of each satellite's orbital angular momentum keeps a small value throughout the history of this halo and (2) the ratio ${\rm c/a}$ of the satellites oscillates around a fairly low value. The reason for point (2) is that angular momentum conservation keeps the satellites moving in their respective orbits while the relative positions of the satellites change periodically thus leading to an oscillation of the ratio ${\rm c/a}$. Because the angular dispersion is small \ (as point (1) shows), the oscillation of the ratio ${\rm c/a}$ is around a small value of $\sim 0.2$ and has a fairly small amplitude  (see \Fg{fig7}). In this case, a rotating thin plane is more likely to be `observed' at present time\ (\ie\ at $z=0$), although the ratio ${\rm c/a}$ is oscillating and today's value is by no means its lowest \ (i.e. at a lookback time of $\sim$ 6 Gyrs, the value of $c/a \approx 0.1$). The reader will note  that the other three satellites just happen to appear at the right positions with the right velocities at $z=0$ (see \Fg{fig3} and the orange curve in the third row of \Fg{fig7}), culminating in a MW-like rotating thin plane\ (with 14 members) at that time. We conclude that, for halo395, the rotating thin plane is transient(as the value of  $c/a$ is oscillating and the appearance of the other three satellites is by chance) and that the accretion along the local super-galactic sheets can indeed lead to a rotating satellite plane. It is worth noting that \citet{Sawala2022Na} proves that the satellite plane of MW is also transient by using EDR3 proper motion data.

\section{Discussion} \label{sec:discuss}

Unlike many previous studies(\eg \ \citet{Pawlowski2018}), we don't account for the effect of obscuration of the host stellar disk here. In these studies, they often use a simple correction when selecting satellites: exclude everything that lies within $|b|\leqslant 12^{\circ}$ where b is the galactocentric latitude.  We don't do the correction for two reasons: 1. As mensioned in \Se{obs}, we assumed the satellites sample with $\Mstar \ge 10^{5} \Msun$ is complete, like many other works. 2. As shown by \citet{Samuel2021}, the correction will contaminate the analysis of the origin of satellite planes, however, which is one of the main goals of this work. \par 
%3. Although the probability of finding satellite plane is increased by correction, this is still rare\ (as suggested by \citet{Samuel2021}), which is consistent with our results.   

The halo mass ${\rm M_{200}=4.9\times 10^{11}\Msun}$ of halo395 may  be a concern, as it is at the lower end of observed data. However as some previous studies \citep{Pawlowski2019apj,Samuel2021} found, there are only weak or no correlations between the satellite plane metrics \ (h,  c/a and orbital dispersion) and host mass. We conclude that this isn't  a  particularly critical  problem. Furthermore if the halo mass is low, its only off by a factor of $<2$.\par

As shown by \citet{Sawala2022Na}, the  metric $c/a$ correlates with the satellites radial distribution, since the metric $c/a$ is sensitive to the radius of  satellites. The more centrally concentrated the satellites(higher Gini coefficient, see \citet{Sawala2022Na}), the lower the $c/a$ value. An alternative radius-independent metric is the reduced inertial tensor method \citep{Bailin2005} $(c/a)_{\rm Red}$. The reduced inertial tensor is computed in the same way as the inertia tensor, except it uses unit position vectors. In this work, we use the common metric $c/a$, as we want to find satellite systems that are the same as MW's in terms of full spatial anisotropy\ (by using $c/a$), not just in terms of angular anisotropy\ (by using $(c/a)_{\rm Red}$). Interestingly, halo395, selected by using $c/a$, possess a similar satellites radial distribution as that of MW (see \Fg{fig8}).  Interestingly, it is found that the $(c/a)_{\rm Red}$ of MW and halo395, at z=0, are 0.342 and 0.319 respectively. This means that either c/a of halo395 indicates that the satellites are indeed in a thin plane. 

It is noted that we use a slightly different metric to measure kinematic coherence\ (\ie\  $k-\Delta_k$ see \Se{method}) when compared with some other works\ (\eg\  \citet{Samuel2021}).  In those works, they use the same formula as us, but they only consider a specific k. Since there is no better motivation for any particular value of k over any other, we consider all integers ranging from 4 to 14\ (\ie\  the full satellite sample size). This means the criteria we used is both more strict and more general, and the result we obtain is more robust.\par

We have shown that halo395 is embedded in a sheet plane \ (\Fg{fig2} and \Fg{fig3}) and the formation of satellite plane of halo395 is related to its large scale environment\ (\ie\ voids and sheet). Moreover, the norm of the satellite plane of halo395  points to the void rightly above the sheet plane\ (can be directly seen by eye from \Fg{fig2} and \Fg{fig3}). Interestingly, \citet{Libeskind2015} showed that MW is embedded in Local Sheet and the norm of its satellite plane also points towards the local void above and below the Local Sheet, which suggest that the scenario of the formation of  the satellite plane found here may also apply to the real MW. 
% {\bfseries However,  \citet{Samuel2021,Garavito2021,Santos-Santos2023ApJ} found that the presence of massive satellites \ (like LMC) will increases the possibility of planarity and enhance the fraction of co-rotating satellites.}  The issue about what is the major cause of the satellite plane of MW requires further study.
\par

It is also worth noting that as angular momentum  is a vector, one may want to distinguish the prograde or retrograde orbits differently, as some previous studies\ \citep{Pawlowski2020} have done. Here we also check the  ${\rm k-\Delta_{k}}$ relation using \eq{eq1} without flipping the angular momentum of retrograde satellites and the result is shown in \Fg{fig6}. It shows that the dispersion ${\rm \Delta_{k}}$ of the model galaxy is larger than that of MW when ${\rm k\ge9}$. This is because some model satellites are prograde and  the rest are retrograde, but all of them are orbiting in a thin plane as shown in the \Fg{fig1} and \Fg{fig4}. Although, halo395 is a bit different from the MW in this context, the mechanism we proposed \ (see \Se{origin})  here may still apply to the satellite plane of MW.  However, some recent studies (\eg\  \cite{Li2021ApJ}) found that most fainter members of the satellite plane of MW are co-orbiting. In this case,  it may be hard to explain it with our scenario alone, but \citet{Santos-Santos2023ApJ} pointed that the late capture of LMC-like object could increase the fraction of co-orbiting satellites (notably, there are no LMC analogs in halo395, see \Se{basicChar}). So, the late capture of massive satellites like LMC may be also needed to explain the asymmetry in the direction of orbits.

Of course, we can not draw as a conclusion that the Satellite Plane Problem is solved, by only considering the plane around the MW, as some other  satellite planes were found around other galaxies like M31 and CenA ; there are hints of satellite planes around more distant galaxies well beyond the Local Group \eg\ M81\citep{Chiboucas2013}, M83\citep{Muller2018}, M101\citep{Muller2017}. So, to truly understand the Satellite Plane Problem requires 
\begin{enumerate}
    \item large statistically significant  observational samples which include more distant galaxies\ (\eg \ \citet{Ibata2014,Mao2021}).
    
    \item  high resolution cosmological simulations with large volumes.
\end{enumerate}

\section{ Summary} \label{sec:summary}

In this work, we studied the Satellite Plane Problem  of the MW, by using numerical data of the newly published simulation TNG50-1 which has high resolution  and large volume at the same time.  We searched the simulation to see if any halo had a  a thin, rotating satellite plane like the MW, and if such halos do exist, how do the planes formed and how long can it last.  Below are our major findings:
\begin{enumerate}
    \item One halo\ (haloID$=395$,z=0)  is similar to the MW in the sense of halo Mass, stellar mass, morphology, environment and the satellites radial distribution and does not suffer from the Satellite Plane Problem (out of 231 candidates) in TNG50-1. Thus the MW-like satellite plane can be found in TNG50-1 simulations based on $\Lambda$CDM model.

    \item The spatially thin and kinematically coherent satellite plane found at z=0 is transient \ (see \Fg{fig7} and \Se{origin}).  This is in agreement with the findings of \citet{Sawala2022Na}. 
    
    \item  The major part (11/14) of the satellite plane of halo395 accreted along the sheets. Thus, for our special case (halo395), the accretion of satellites along the sheets leads the formation of a thin and rotating plane\ (see \Fg{fig2}, \Fg{fig3} and \Se{origin}). However, large samples of MW-like halos in different environments are needed to prove whether a special geometry (\ie \ sheet and voids ) can boost the formation of  thin and rotating plane, which is beyond the scope of our current work. 

    % \item {\bfseries  In addition to the mechanism we proposed, the late capture of massive satellites like LMC may also be needed to account for the observation: most members of the satellite plane of MW are orbiting in the same sense.(see \Se{discuss}).}
\end{enumerate}

\begin{acknowledgments}
We thank the anonymous referee for very useful comments. We thank  Marcel S. Pawlowski for his helpful suggestions and comments. We are also grateful to Yaoxin Chen for nice discussions. Yingzhong Xu and Xi Kang acknowledge the support from the National Key Research and Development Program of China (No.2022YFA1602903), the NSFC \ (No. 11825303, 11861131006), the science research grants from the China Manned Space project with No. CMS-CSST-2021-A03, CMS-CSST-2021-A04, the Fundamental Research Funds for the Central Universities of China \ (226-2022-00216) and the start-up funding of Zhejiang University.  
\end{acknowledgments}


% \vspace{5mm}
% \facilities{HST(STIS), Swift(XRT and UVOT), AAVSO, CTIO:1.3m,


\vspace{5mm}
\software{Astropy \ \citep{Astropy2013,Astropy2018},
Numpy \ \citep{Walt2011,Harris2020}, Scipy\ \citep{Virtanen2020}, Matplotlib\ \citep{Hunter2007} 
}



% \appendix

% \section{Appendix information}


\bibliography{reference}{}
\bibliographystyle{aasjournal}

\end{document}